\documentclass[cernpreprint, atlasdraft=false, texlive=2023, UKenglish, texmf, orcidlogo]{atlasdoc}
 
\usepackage[subfigure,rotating]{atlaspackage}
\usepackage{atlasbiblatex}
 
\usepackage{atlasphysics}

\graphicspath{{logos/}{figures/}}
 
\usepackage{ANA-EXOT-2020-06-PAPER-defs}

\AtlasTitle{Search for Majorana neutrinos in same-sign $WW$ scattering events from $pp$ collisions at $\sqrt{s}=13$~TeV}

\AtlasAbstract{
A search for Majorana neutrinos in same-sign $WW$ scattering events is presented. The analysis uses $\sqrt{s}= 13$~TeV proton--proton collision data with an integrated luminosity of 140~fb$^{-1}$ recorded during 2015--2018 by the ATLAS detector at the Large Hadron Collider. The analysis targets final states including exactly two same-sign muons and at least two hadronic jets well separated in rapidity. The modelling of the main backgrounds, from Standard Model same-sign $WW$ scattering and $WZ$ production, is constrained with data in dedicated signal-depleted control regions. The distribution of the transverse momentum of the second-hardest muon is used to search for signals originating from a heavy Majorana neutrino with a mass between 50~GeV and 20~TeV. No significant excess is observed over the background expectation. The results are interpreted in a benchmark scenario of the Phenomenological Type-I Seesaw model. In addition, the sensitivity to the Weinberg operator is investigated. Upper limits at the  95\% confidence level are placed on the squared muon-neutrino--heavy-neutrino mass-mixing matrix element $\vert V_{\mu N} \vert^{2}$ as a function of the heavy Majorana neutrino's mass $m_N$, and on the effective $\mu\mu$ Majorana neutrino mass $|m_{\mu\mu}|$. 
}

\AtlasRefCode{EXOT-2020-06}
 
\PreprintIdNumber{CERN-EP-2023-099}

\AtlasJournal{EPJC}
\AtlasJournalRef{Eur. Phys. J. C 83 (2023) 824}
\AtlasDOI{10.1140/epjc/s10052-023-11915-y}

\hypersetup{pdftitle={ATLAS document},pdfauthor={The ATLAS Collaboration}}
 
\begin{document}
 
\maketitle

\section{Introduction}
 
The Higgs field's Yukawa couplings explain in a satisfactory way, and in agreement with data, the generation of the charged-fermion masses. However, the precise origin of tiny but non-zero neutrino masses, brought to light by neutrino oscillation experiments~\cite{Super-Kamiokande:2005mbp, SNO:2002tuh}, is not explained by the Standard Model (SM) of particle physics and remains a mystery today. Among the most compelling solutions to this mystery is the hypothesised existence of Majorana masses for neutrinos, which can be explored by high- and low-energy experiments~\cite{Atre:2009rg,Deppisch:2015qwa,Cai:2017jrq,Cai:2017mow}.
 
Majorana neutrino masses are present in beyond-the-SM (BSM) theories that include new, heavy leptons or extended scalar sectors, collectively known as Seesaw models, or in theories with extended gauge sectors (Left-Right Symmetric Model, Grand Unified Theories)~\cite{Ma:1998dn}. Even if these new particles have masses that are much higher than the energies reached at the Large Hadron Collider (LHC), and are therefore kinematically inaccessible, the Majorana nature of active neutrinos can still be probed indirectly using effective field theories, with the lowest-dimensional operator being the so-called dimension $d=5$ Weinberg operator~\cite{Weinberg:1979sa}.

\begin{figure}[h]
\label{fig:ssWWfeyn}
\begin{center}
\subfigure[]{\includegraphics[width=0.45\linewidth]{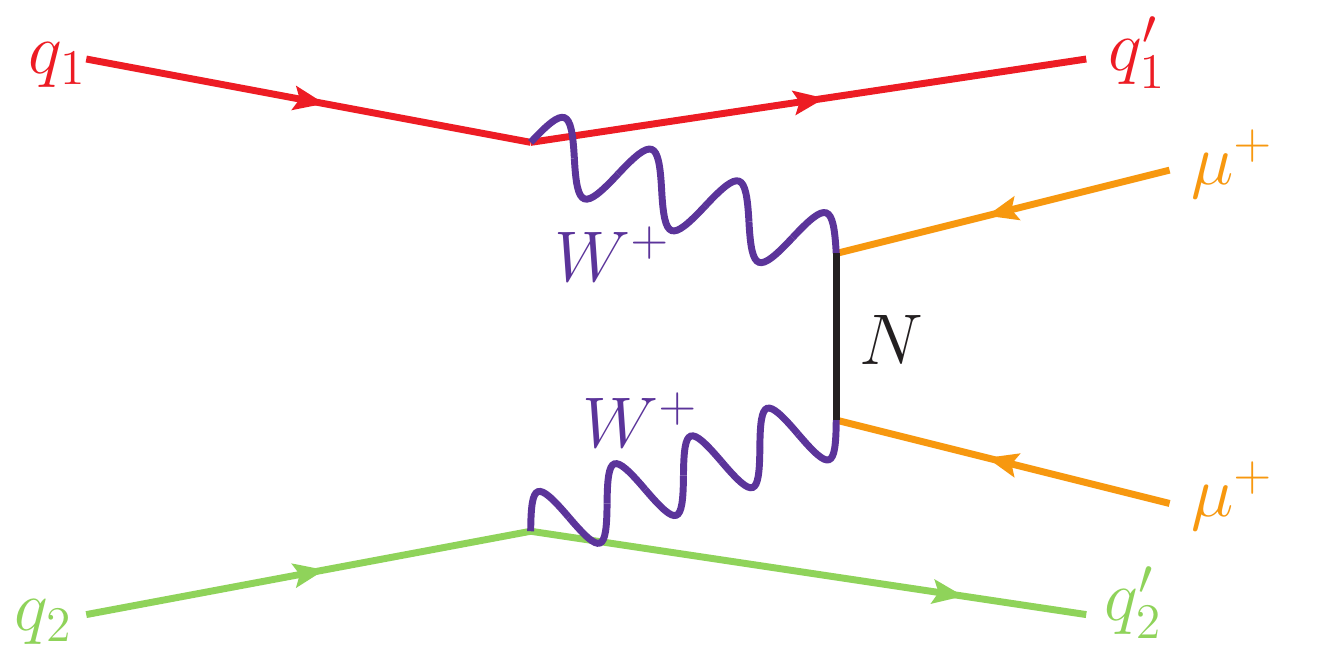}}
\subfigure[]{\includegraphics[width=0.45\linewidth]{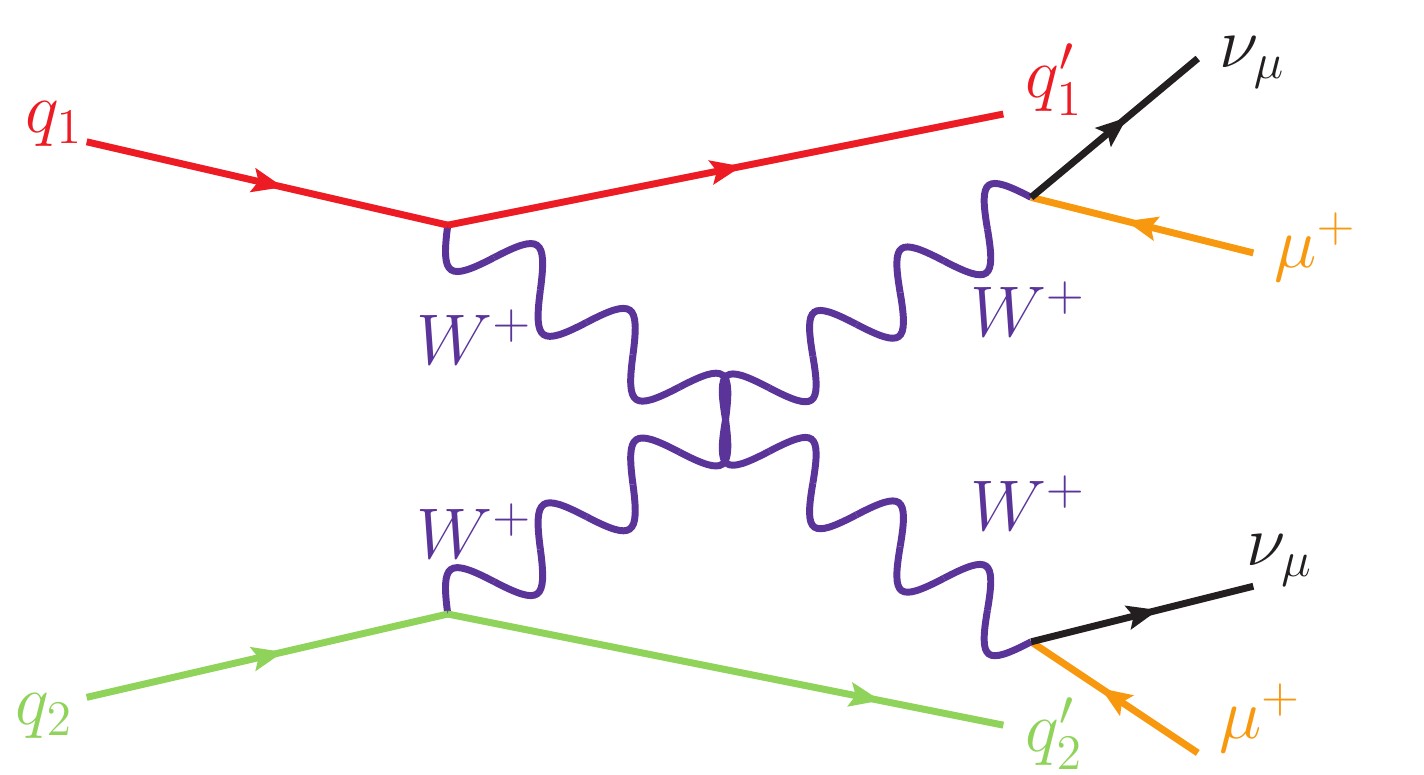}}
\caption{
Diagrammatic representation of (a) same-sign $\mu^{+}\mu^{+}$ production in $W^{+} W^{+}$ scattering mediated by a Majorana neutrino $N$ in proton--proton collisions and (b) electroweak same-sign $W^{+}W^{+}$ scattering, which is the main background in this search. The corresponding diagrams with negative-charged particles are also relevant to this work but not included here for simplicity.
}
\end{center}
\end{figure}

At energies below 100~\MeV, the Majorana nature of neutrinos can be probed by searching for neutrinoless double beta decays of certain isotopes, which are characterised by the appearance of two same-sign electrons ($e$) with an absence of neutrinos ($\nu$) in the final state~\cite{Dolinski:2019nrj}. Although the production of same-sign leptons ($\ell$) involving muons ($\mu$) or $\tau$-leptons is kinematically forbidden in such isotope decays, the same process can be studied at high energies when mediated at dimension
$d=5$, 7, or higher
~\cite{Fuks:2020att,Fuks:2020zbm, Dicus:1991fk}, as shown in Figure~\ref{fig:ssWWfeyn}(a). This is the subject of the present work, which searches for signs of heavy neutral leptons in the same-sign dimuon final state with the ATLAS detector. This search also connects to existing LHC searches for heavy Majorana neutrinos in resonant production~\cite{ATLAS:2019kpx,ATLAS:2015gtp,CMS-EXO-17-028,CMS:2018iaf}, with the goal of extending the reach for heavy neutrino masses to beyond the \TeV~scale~\cite{Fuks:2020att}, where the resonant production channels become kinematically inaccessible. A similar search was reported recently by the CMS Collaboration~\cite{CMS-EXO-21-003}. Searches for heavy Majorana neutrinos lighter than the $Z$ boson were conducted at LEP~\cite{DELPHI:1996qcc}.

This search is conducted in the context of two benchmark models. The first is the Phenomenological Type-I Seesaw model~\cite{delAguila:2008cj,Atre:2009rg}, where heavy Majorana neutrinos couple to SM particles through mass-mixing with light neutrinos, which is parameterised with a complex-valued matrix element. For simplicity, this work investigates only the phenomenology of the lightest heavy-neutrino mass eigenstate $N$ by assuming that all other heavy eigenstates are decoupled. 
The benchmark scenario is considered without $e$ and $\tau$ flavour mixing, so the only non-zero entry in the mass-mixing matrix is the element $\vert V_{\mu N}\vert^2$. In this case the signal cross section $\sigma(pp \to \mu^\pm \mu^\pm+X)$ can be written as a function of the cross section $\sigma_0(pp \to \mu^\pm \mu^\pm+X)$ for the flavour-aligned scenario  (in which $|V_{\mu N}|=1$):
\begin{equation}
\sigma(pp \to \mu^\pm \mu^\pm+X) \equiv \vert V_{\mu N}\vert^4 \times \sigma_0(pp \to \mu^\pm \mu^\pm+X).
\label{eq:bareXSecT}
\end{equation}
The two free parameters of the model are $\vert V_{\mu N}\vert^2$ and the mass $m_N$ of the heavy Majorana neutrino.
 
The second benchmark model is the $d=5$ Weinberg operator~\cite{Weinberg:1979sa}, which extends the SM Lagrangian by:
 
\begin{equation*}
\mathcal{L}_{5} = \ \frac{C_5^{\ell\ell'}}{\Lambda} \big[\Phi\!\cdot\! \overline{L}^c_{\ell }\big]
\big[L_{\ell'}\!\!\cdot\!\Phi\big], 
\end{equation*}
 
where $\Lambda$ is the scale at which the particles mediating the BSM signal become relevant degrees of freedom; $C_5^{\ell\ell'}$ is a complex, flavour-dependent Wilson matrix coefficient; $L_\ell^T=(\nu_\ell,\ell)$ is the left-handed lepton doublet; and $\Phi$ is the SM Higgs doublet. Considering the Higgs vacuum expectation value $\upsilon=\sqrt{2}\langle\Phi\rangle\approx246~\GeV$, it is possible to define the quantity
 
\begin{equation}
m_{\ell\ell'} = C_5^{\ell\ell'} \upsilon^2/\Lambda,
\label{eq:dim5Mass}\end{equation}
 
which for $\ell \ell' = ee$ reduces to the effective Majorana mass usually quoted by neutrinoless double beta decay experiments, which is a measure of the scale at which lepton-number symmetry is broken~\cite{Dolinski:2019nrj, Fuks:2020zbm}. For simplicity, it is assumed here that $C_5^{\mu\mu}$ is the only non-zero Wilson coefficient of this matrix. The signal cross section $\sigma(pp \to \mu^\pm \mu^\pm+X)$ can then be written as a function of the dimensionless effective cross section $\sigma_5(pp \to \mu^\pm \mu^\pm+X)$ for
$\Lambda/|C_5^{\mu\mu}|=1~\TeV$ as follows:
 
\begin{equation}
\sigma(pp \to \mu^\pm \mu^\pm+X) \equiv \frac{|C_5^{\mu\mu}|^2}{\Lambda^2} \cdot \sigma_5(pp \to \mu^\pm \mu^\pm+X) \, .
\label{eq:bareXSecT_weinberg}
\end{equation}
 
The analysis targets final states with two same-sign muons, and also two particle jets with a large angular separation to capture the signature of vector-boson scattering. 
The main backgrounds are SM $\sswwjj$ production, shown in Figure~\ref{fig:ssWWfeyn}(b), and $WZ$ production, where the bosons decay leptonically. They are estimated with simulated events and their modelling is constrained in dedicated signal-depleted control regions. A data-driven technique assisted by simulation is used to
model backgrounds which do not originate from prompt muons.
The presence of a signal is tested in a region designed to be signal-enriched
using the benchmark models discussed above.


\section{The ATLAS detector}
 
The ATLAS detector~\cite{PERF-2007-01} at the LHC is a multipurpose particle detector
with a forward--backward symmetric cylindrical geometry with a nearly \(4\pi\) coverage in
solid angle.\footnote{ATLAS uses a right-handed coordinate system with its origin at the nominal interaction point (IP) in the centre of the detector and the \(z\)-axis along the beam pipe. The \(x\)-axis points from the IP to the centre of the LHC ring, and the \(y\)-axis points upwards. Cylindrical coordinates \((r,\phi)\) are used in the transverse plane, \(\phi\) being the azimuthal angle around the \(z\)-axis. The pseudorapidity is defined in terms of the polar angle \(\theta\) as \(\eta = -\ln \tan(\theta/2)\).
}
It consists of an inner tracking detector (ID) surrounded by a superconducting solenoid,
electromagnetic (EM) and hadron calorimeters, and a muon spectrometer (MS).
 
The ID covers the pseudorapidity range \(|\eta| < 2.5\).
The high-granularity silicon pixel detector covers the vertex region and typically provides four measurements per track, the first hit normally being in the insertable B-layer installed before Run 2~\cite{ATLAS-TDR-19, PIX-2018-001}.
It is surrounded by a silicon microstrip detector and a straw tube transition-radiation tracking detector.
The calorimeter system covers the pseudorapidity range \(|\eta| < 4.9\).
It consists of lead/liquid-argon (LAr) sampling calorimeters which provide EM energy measurements
with high granularity up to \(|\eta| = 3.2\).
A steel/scintillator-tile hadron calorimeter covers the central pseudorapidity range (\(|\eta| < 1.7\)).
The endcap and forward regions are instrumented with LAr calorimeters
for EM and hadronic energy measurements up to \(|\eta| = 4.9\).
The MS surrounds the calorimeters and is based on
three large air-core toroidal superconducting magnets with eight coils each.
The field integral of the toroids ranges between \num{2.0} and \SI{6.0}{\tesla\metre} across most of the detector.
The MS includes a system of precision chambers for tracking and fast detectors for triggering.
The ATLAS trigger and data acquisition system~\cite{TRIG-2016-01} contains a two-level trigger system in order to select events of interest.
The first-level trigger is implemented in hardware and uses a subset of the detector information
to accept events at a maximum rate of nearly \SI{100}{\kHz}.
The second-level is a software-based trigger that
reduces the accepted event rate to \SI{1}{\kHz}, on average,
depending on the data-taking conditions~\cite{TRIG-2016-01}.
An extensive software suite~\cite{ATL-SOFT-PUB-2021-001} is used in data simulation, in the reconstruction
and analysis of real and simulated data, in detector operations, and in the trigger and data acquisition systems of the experiment.


\section{Object reconstruction and event selection}
\label{sec:selection}
 
This analysis uses a set of $pp$ collision data events collected by the ATLAS detector between 2015 and 2018 at a centre-of-mass energy of
$\sqrt{s}=13$~\tev.
Only events for which all detector subsystems were operational are considered.
The data set corresponds to an integrated luminosity of 140~\ifb~\cite{DAPR-2021-01,LUCID2}.
 
This analysis is based on events where the detector readout is triggered by the presence of at least one muon, referred to as \enquote{single-muon triggers}. Single-muon triggers require the presence of a single muon with a minimum transverse momentum (\pt) in the range
20--26~\GeV\ depending on the data-taking period~\cite{TRIG-2018-01}.
 
Events are required to have at least one collision vertex reconstructed from at least two ID tracks with
\(\pt > 500~\MeV\). For events with several collision vertices, the one with the largest sum of the squared
transverse momenta of the associated tracks is taken as the primary vertex~\cite{ATL-PHYS-PUB-2015-026}.

Three types of muon candidates are used to select events for the signal region, to estimate the background contribution from fake muons, and to veto additional muons in the event selection. These muon candidates have different selection criteria and are referred to as  \enquote{signal muons}, \enquote{spurious muons}, and \enquote{baseline muons}, respectively. Baseline muon candidates are reconstructed by combining tracks in the ID with tracks in the MS~\cite{PERF-2015-10} and are required to have $|\eta|<2.5$ and $\pt > 3~\GeV$. Further selections on the longitudinal and transverse impact parameters
are imposed. The transverse impact parameter $d_0$, measured from the beamline and divided by its estimated uncertainty, must satisfy ${|\dzero|/\sigma(\dzero)} < 15$. The longitudinal impact parameter $z_0$, measured from the primary vertex,
is required to satisfy ${|\zzsth|<1.5}$~mm. Baseline muon candidates must satisfy the \enquote{Loose} cut-based identification working point defined in Ref.~\cite{MUON-2018-03}, based on the numbers of hits in the different ID and MS subsystems, and on the significance of the charge-to-momentum ratio $q/p$. To select signal muons, these selection criteria are tightened to $\pt > 10$~\GeV, $|\eta|<2.5$, ${|\dzero|/\sigma(\dzero)}<3$ and ${|\zzsth|<0.5}$~mm, and used together with the \enquote{Medium} cut-based identification working point. In addition, an isolation is required where the scalar sum of the $\pt$ of tracks inside a variable-size cone around the muon (excluding its own track) plus 40\% of the transverse energy of neutral particle-flow objects in a cone of size $\Delta R = 0.2$ around the muon, must be less than 4.5\% of the muon $\pt$~\cite{MUON-2018-03}. To select spurious muons, the above isolation requirement is loosened to be less than 16\% of the muon $\pt$ and ${|\dzero|/\sigma(\dzero)}<10$. In addition, they must fail the tighter signal-muon isolation or transverse impact parameter requirements.

Electron candidates are reconstructed from an isolated EM calorimeter
energy deposit which is matched to a track in the ID~\cite{EGAM-2018-01} and they are required to have $\pt > 4.5~\GeV$.
The pseudorapidity of the calorimeter energy cluster, $\eta_{\text{cluster}}$, must satisfy $|\eta_{\text{cluster}}|<2.47$. 
Electron candidates must pass a \enquote{Loose} likelihood-based identification working point~\cite{EGAM-2018-01}
that employs calorimeter and tracking information to discriminate between electrons and jets, and their tracks must have
impact parameters satisfying ${|\dzero|/\sigma(\dzero)} < 5$ and ${|\zzsth|<0.5}$~mm. In this analysis, electron candidates are used only to veto events.
 
Jets are reconstructed by using the \antikt algorithm~\cite{Cacciari:2008gp,Fastjet} with a radius parameter of \(R=0.4\)
after applying a particle-flow algorithm~\cite{PERF-2015-09} to tracking information and topological clusters of calorimeter-cell energies.
Jets are required to have \(\pT > 25~\GeV\) and \(|\eta| < 4.5\), and
are calibrated as described in Ref.~\cite{JETM-2018-05}.
The jet-vertex tagger (JVT) discriminant \cite{ATLAS-CONF-2014-018} is used to identify
jets originating from the hard-scatter interaction through the use of tracking and
vertexing information. Jets with \(\pT < 60~\GeV\) and
\(|\eta| < 2.5\) are required to satisfy the requirement \(\text{JVT} > 0.50\),
corresponding to a selection efficiency of about \SI{96}{\percent} for hard-scatter jets~\cite{PERF-2014-03}.
If either of the two leading jets has $|\eta|{\,>\,}2.5$ and \(\pT < 60~\GeV\), it must satisfy a threshold for the forward jet-vertex tagger (fJVT) discriminant~\cite{ATL-PHYS-PUB-2015-034,ATL-PHYS-PUB-2019-026} that has an efficiency of about 93\% for hard-scatter jets.
 
In order to suppress contributions from background processes that involve top quarks or leptonic $b$-hadron decays, the DL1r classification algorithm based on recurrent neural networks~\cite{FTAG-2019-07,
ATLAS-CONF-2018-001} is used to identify jets containing $b$-hadrons (referred to as \enquote{\(b\)-jets}).
This algorithm identifies \(b\)-jets amongst a background of light-flavour- and charm-quark-initiated
jets using information about the impact parameters of tracks associated with the jet
and the topological properties of displaced vertices reconstructed within the jet.
This analysis uses a \(b\)-jet efficiency working point of \SI{85}{\percent} as measured
for jets with \(\pT > 20~\GeV\) and $|\eta| < 2.5$ in simulated \ttbar events~\cite{FTAG-2018-01}.
The tagging algorithm at that working point gives an expected rejection factor (defined as the inverse of the efficiency) of about 40 for light-flavour jets,
and about 2.9 for jets originating from charm quarks~\cite{ATLAS-CONF-2018-001,ATLAS-CONF-2018-006}.
 
To resolve the potential ambiguities due to a single detector response being assigned to two
objects by the reconstruction algorithm, a sequential overlap-removal procedure is applied.
First, any electron found to share a track with another electron with higher \pT is removed. Second, electrons sharing their track with a muon candidate are removed. Third, labelling the rapidity of an object by $y$,\footnote{The rapidity is defined as $y=\frac{1}{2} \ln{\frac{E+p_z}{E-p_z}}$,
where $E$ is the energy and $p_z$ is the longitudinal component of the momentum along the beam pipe.} any jet within $\Delta R_y = \sqrt{(\Delta y)^2+(\Delta \phi)^2} = 0.2$ of
an electron is removed, unless it is a $b$-jet and the electron \pt\ is below 100~\GeV, in which case the electron is removed.
Fourth, any electrons within $\Delta R_y = 0.4$ of a remaining jet are removed. Fifth, any jet that is within $\Delta R_y = 0.2$ of a muon and has less than three associated tracks is removed, unless it is a $b$-jet, in which case the muon is removed.
Finally, remaining muons are removed if their track is within $\Delta R_y = 0.4$ 
of a remaining jet. Resolving the overlaps between $e/\mu$ and jets in two steps based on $\Delta R_y$ values of 0.2 and 0.4 allows the reconstruction efficiency to be maximised for $b$-jets including semileptonic $c$- and $b$-hadron decays.
 
The missing transverse momentum, with magnitude $\MET$, is defined as the negative vectorial sum of the transverse momenta of all selected and calibrated electrons, muons and jets, and an additional term to account for soft energy in the event that is not associated with any of the
selected objects. This soft term is calculated from unassociated ID tracks matched to the primary vertex. While it suffers from an inherent underestimation, which is covered by dedicated uncertainties, it is more resilient to contamination from multiple interactions in the same or neighbouring bunch crossings (pile-up) compared to the calorimeter-based soft term~\cite{PERF-2016-07,ATLAS-CONF-2018-023}.
The object-based \met significance ($\mathcal{S}$) is computed~\cite{ATLAS-CONF-2018-038} by including the expected resolutions for all the objects used in the \met calculation and is used to reject background events with genuine \met\ originating from undetected particles such as neutrinos.

At preselection level, the events are required to have at least two same-sign signal muons with $\pT>\SI{27}{\GeV}$ and one of them has to
match the muon that fired the trigger. At least two jets are required and one of the jets has to have a transverse momentum larger than 30 GeV to reduce the contamination from \pileup jets. Events with $b$-jets are vetoed.
In order to preferentially capture the signature of $W^{\pm}W^{\pm}$ scattering in $pp$ collisions,
the invariant mass of the two jets with the highest \pt,
labelled as $m_{jj}$, has to be larger than 300~\GeV\ and their rapidity separation must satisfy $|\Delta y_{jj}| > 4$. These preselection cuts are applied to every event which is used further in the analysis.


\section{Event simulation}
 
Monte Carlo (MC) samples are used to model the expected signal and background yields, while backgrounds from misidentified objects are estimated using data-driven techniques described in
Section~\ref{sec:analysis}.
All MC samples
are normalised to the best theory predictions available.
The \NNPDF[3.0nlo]~\cite{Ball:2014uwa} PDF set was used in all
matrix element calculations if not stated otherwise.
The generated event samples were processed through the full ATLAS detector simulation~\cite{SOFT-2010-01}
based on \GEANT~\cite{Agostinelli:2002hh}. Additional simulated \pp collisions generated
using \PYTHIA[8.186]~\cite{Sjostrand:2007gs} with the \textsc{A3} set of tuned parameters (tune)~\cite{ATL-PHYS-PUB-2012-003}
and the \MSTW[lo] parton distribution function (PDF) set~\cite{Martin:2009iq} were overlaid to model the effects of \pileup. The distribution of the number of
additional \pp interactions in the MC samples is reweighted to match the one observed in data. All
simulated samples are processed through the same reconstruction algorithms and analysis chain as the data.
 
The Phenomenological Type-I Seesaw model signal samples were simulated following the prescription given in Ref.~\cite{Fuks:2020att}.
The samples were simulated using the \MGNLO[2.7.2]~\cite{Alwall:2014hca} generator at next-to-leading order (NLO) in QCD after importing the default variant of the \textsc{HeavyN}~\cite{Degrande:2016aje, Alva:2014gxa} \textsc{FeynRules} UFO libraries~\cite{Christensen:2009jx, Alloul:2013bka}. All \textsc{HeavyN} events were interfaced to \PYTHIA[8.243]~\cite{Sjostrand:2014zea}. Twenty mass points were generated, spanning the range $m_{N} = 50$\,\GeV~to 20\,\TeV. Since the mixing parameter $V_{\mu N}$ only impacts the total cross section, as $\sigma\sim |V_{\mu N}|^{4}$, these mass points were generated with a mixing parameter of unity and are rescaled to other values by using Eq.~(\ref{eq:bareXSecT}).
 
The signal sample for the Weinberg operator was simulated by using the prescription given in Ref.~\cite{Fuks:2020zbm}. For the hard parton-level scattering processes, \MGNLO[2.9.3]~\cite{Alwall:2014hca} was used at NLO in QCD after importing the default variant of the \textsc{SMWeinberg}~\cite{Fuks:2020zbm} \textsc{FeynRules} UFO libraries~\cite{Christensen:2009jx, Alloul:2013bka}. The NNPDF3.1luxQED~\cite{Ball:2014uwa} PDF was used. All \textsc{SMWeinberg} events were interfaced to \PYTHIA[8.245]~\cite{Sjostrand:2014zea}.
The sample was generated at an energy scale $\Lambda=200$~\TeV\ and with the Wilson coefficient value $C_5^{\mu\mu}=1$. Both $\Lambda$ and $C_5^{\mu\mu}$ only impact the total cross section, as $\sigma\sim (C_5^{\mu\mu}/\Lambda)^2$, so a global rescaling of a single signal sample according to Eq.~(\ref{eq:bareXSecT_weinberg}) is used to vary their values.
 
For all samples of simulated signal events, the parton shower, hadronisation, and underlying-event modelling by \PYTHIA used the A14 tune~\cite{ATL-PHYS-PUB-2014-021} and the \NNPDF[2.3lo]~\cite{Ball:2014uwa} PDF set, and the \EVTGEN[1.7.0]
program~\cite{Lange:2001uf} was used to model the decays of bottom and charm hadrons.
 
Electroweak (EW) $W^\pm W^\pm$ production and decay in association with two jets was simulated with diagrams including exactly six orders of the EW coupling~\cite{ATL-PHYS-PUB-2019-004}. The simulation of the strong production processes, and the interference between EW and strong contributions, includes diagrams with exactly four and five EW vertices~\cite{ATL-PHYS-PUB-2017-005}, respectively. All SM $\sswwjj$ contributions were simulated with \MGNLO[2.6.7]~\cite{Alwall:2014hca}, which was interfaced to the \PYTHIA[8.244]~\cite{Sjostrand:2014zea} parton shower model. The matching of the matrix element simulation to parton showers used the MC@NLO method~\cite{Frixione:2002ik} based on FKS subtraction~\cite{Frederix:2009yq}. These samples used the A14 set of tunable parameters~\cite{ATL-PHYS-PUB-2014-021}. More details of the generator settings used to obtain a better description of deep-inelastic scattering data~\cite{Cabouat:2017rzi} are given in Ref.~\cite{ATL-PHYS-PUB-2019-004}.
 
Samples of diboson final states ($VV$, $V=W/Z$) were simulated with the
\SHERPA[2.2.1] or 2.2.2~\cite{Bothmann:2019yzt} generator depending on the process.
Fully leptonic final states and semileptonic final states, where one boson
decays leptonically and the other hadronically, were generated using
matrix elements with NLO accuracy in QCD for up to one additional parton
and with LO accuracy for up to three additional parton
emissions. Samples for the loop-induced processes $gg \to VV$ were
generated using LO matrix elements for up to one
additional parton emission.
The production of triboson ($VVV$) events was simulated with the
\SHERPA[2.2.2] generator, accurate to NLO in QCD for the inclusive process and to LO for up to
two additional parton emissions, using factorised gauge-boson decays.
The matrix element calculations for individual $VV$ and $VVV$ processes were matched
and merged with the \SHERPA parton shower based on Catani--Seymour
dipole factorisation~\cite{Gleisberg:2008fv,Schumann:2007mg} using the MEPS@NLO
prescription~\cite{Hoeche:2011fd,Hoeche:2012yf,Catani:2001cc,Hoeche:2009rj}.
The virtual QCD corrections for matrix elements with NLO accuracy were
provided by the \OPENLOOPS[1] library~\cite{Buccioni:2019sur,Cascioli:2011va,Denner:2016kdg}.
 
Additional samples are used to simulate minor backgrounds with prompt muons.
The production of \ttV , $t\bar{t}WW$, \tZq\ and \tWZ\ events was modelled using the
\MGNLO[2.3.3] generator at NLO, interfaced to \PYTHIA~\cite{Sjostrand:2014zea}. The \SHERPA[2.2.11] generator was used to simulate $V\gamma$ final states.
For studies of background which does not originate from prompt muons, the production of $t\bar{t}$ and single-top-quark events was simulated using the
\POWHEGBOX[v2]~\cite{Frederix:2012dh,Nason:2004rx,Frixione:2007vw,Alioli:2010xd}
generator at NLO in QCD, interfaced to \PYTHIA[8.230]~\cite{Sjostrand:2014zea}, and
the production of $V+$jets events was simulated with the \SHERPA[2.2.11] generator.


\section{Analysis strategy}
\label{sec:analysis}
 
The signal considered in this search is characterised by two same-sign muons with large transverse momentum and two jets with a large rapidity separation, which is already exploited in the preselection requirements. The signal region (SR) was optimised for the Phenomenological Type-I Seesaw model with Majorana neutrino masses above 1~\TeV, since this is the kinematic region where this search is
expected to bring unique sensitivity. The best discrimination between signal and background  is given by the distribution of the transverse momentum of the second-highest-\pt\ muon  (\pTmu ). The distribution has a maximum above 200~\GeV\ for $m_N=500$~\GeV, with a long tail towards higher momenta~\cite{Fuks:2020att}, while all SM backgrounds have a steeply falling \pTmu distribution. The main background contributions arise from SM $\sswwjj$ and $WZ$ production.
The SR is defined by requiring the \met to have a small significance value $\mathcal{S} < 4.5$, and events with a third baseline muon or electron are vetoed. This suppresses background from SM $\sswwjj$ events, which have two neutrinos in the final state (cf.\ Figure~\ref{fig:ssWWfeyn}(b)), and $WZ$ events, which have a third lepton.
In contrast, these two requirements are inverted to define signal-depleted control regions (CR) where the data are used to constrain the normalisation of these simulated background processes. The event selection for the SR and the two CRs is summarised in Table~\ref{tab:cuts-sr-crdilep-cr3l}.
 
A same-sign $\sswwjj$ control region (ssWW-CR) is selected with $\mathcal{S} > 5.8$. In order to suppress possible contamination from signal events,  \pTmu is restricted to be below 120~\GeV. This event selection also ensures that background from misidentified jets is adequately suppressed. The ssWW-CR has a SM $\sswwjj$ purity of 70\% inclusively (64\% for $80<\pTmu <120$~GeV), with the main contamination coming from $WZ$ events, whose modelling is, however, constrained in another dedicated control region.
Events which fulfil the ssWW-CR requirement but with a \met significance $4.5 < \mathcal{S} < 5.8$ or with $\pTmu >120$~\GeV\ could have a non-negligible signal contribution. These two regions are not included in the profile likelihood fit, but are used to validate the modelling of SM $\sswwjj$ and $WZ$ events. The final background predictions and the collision data agree well in both validation regions.
 
The second-largest background in the SR is $WZ$ production where both bosons decay leptonically and the $Z$-boson-decay lepton opposite in sign to the $W$ boson falls outside the detector acceptance (mainly due to the muon spectrometer's rapidity coverage). To select a control region enriched in $WZ$ production (WZ-CR), the SR is modified such  that events are required to have three signal muons, with one opposite-sign (OS) muon pair to be compatible with a $Z$ boson decay. Events with an additional baseline muon are vetoed to reduce the contribution from $ZZ$ production. Furthermore, a large three-lepton invariant mass ($m_{\ell\ell\ell} > 100$~\GeV) is required in order to reduce the fake-lepton contribution and to make the WZ-CR orthogonal to the regions where fake leptons from photons are studied, as explained later in this section. The WZ-CR has a $WZ$ purity of 88\% inclusively and 83\% for $\pTmu > 120$~GeV.

\begin{table}[t]
\centering
\caption{
Summary of the main kinematic requirements for the different selection regions. The first four requirements belong to the preselection stage and are described in Section~\ref{sec:selection}.}
\label{tab:cuts-sr-crdilep-cr3l}
\begin{tabular}{l|l|l|l}
\hline\hline
Observable             & SR                        & ssWW-CR        & WZ-CR \\ \hline
Same-sign muons & \multicolumn{3}{c}{$=2$ (signal $\mu$)}\\
Number of $b$-jets & \multicolumn{3}{c}{$=0$}\\
$m_{jj}$   & \multicolumn{3}{c}{$> 300\,\GeV$}\\
$|\Delta y_{jj}|$   & \multicolumn{3}{c}{$>4$}\\ \hline
Third lepton (OS)         & $=0$ (baseline)    & $=0$ (baseline)                          & $=1$ (signal $\mu$)\\
$\met$ signif. $\mathcal{S}$          & $< 4.5$          & $> 5.8 $                                 & $<4.5$ \\
$m_{\ell\ell\ell}$              & ---                         & ---                                      & $ > 100\,\GeV$\\
$p_\mathrm{T}^{\mu_2}$ & ---                        & $<120\,\GeV$  & --- \\[0.2em]
\hline\hline
\end{tabular}
\end{table}
 
Non-prompt muons, defined as originating from bottom- or charm-hadron, kaon or pion decays,
constitute a non-negligible source of background in the analysis. Most of these muons arise from $W$+jets events and $t\bar{t}$ events where one $W$ boson (originating from a top-quark decay in the $t\bar{t}$ case) decays leptonically. This background contribution is evaluated using the so-called fake-factor method~\cite{EGAM-2019-01}, which is based on collision data. Fake factors are derived as a function of muon \pt\ in a dedicated region enriched in non-prompt muons which selects events that include a jet recoiling almost back-to-back against a fake muon candidate. In particular, the presence of $W$+jets events in this sample leads to a non-negligible contamination from prompt muons. This contamination is reduced by requiring the transverse mass formed by the non-prompt muon candidate's \pt\ and the missing transverse momentum to be below 50~\GeV. The residual prompt-muon contribution, estimated from MC predictions, is subtracted from the data,
and the fake factors are calculated as the ratio of non-prompt \enquote{signal} to \enquote{spurious} muon events. The background from non-prompt muons is then estimated by multiplying the fake factors by the data yields observed in the so-called \enquote{spurious-SR}, which has the exact same requirements as the SR except that the object definition of one of the muons is changed from \enquote{signal} to \enquote{spurious}. Non-prompt contributions originating from either the leading or subleading muon candidates are considered. The contamination from prompt muons in the spurious-SR, originating mostly from same-sign $WW$ and $WZ$ events, is subtracted using MC predictions. The final data-driven estimate gives a low non-prompt-muon event yield in the different analysis regions, and hence this contribution to the \pTmu distribution is smoothed by fitting it with an exponential function. Alternative functions are used to estimate the systematic uncertainty due to the functional form choice.
 
Background from \ttV , $t\bar{t}WW$, \tWZ, \tZq, $ZZ$ and $V\gamma$ events is estimated from MC simulation and denoted as \enquote{Other} backgrounds in the remainder of the paper, in which $ZZ$ and \tZq constitute the dominant contributions.
 
Photons originating from final-state radiation splitting into muons ($\gamma^*\to\mu\mu$) were studied in a region selecting leptonic $Z$ boson decays produced in association with an extra muon (which is the non-prompt candidate from final-state radiation) with $m_{\ell\ell\ell}<100$~\GeV\ to suppress the events with initial-state radiation. Their contribution to the SR was found to be negligible. Muons with the wrong reconstructed charge were studied with MC simulation and were also found to be negligible. The probability of misreconstructing two inelastic collisions, each contributing one of a pair of same-sign $W$ bosons, as a single event was evaluated. A Gaussian function with a standard deviation as measured in Ref.~\cite{ATLAS-CONF-2010-027} was used to model the longitudinal distribution of inelastic collisions and to compute the conditional probability. Less than 0.1 events from this source are expected to pass the analysis selection. Finally, single $pp$ interactions where more than one set of partons from each proton interact (double parton scattering) to produce two $W$ bosons with the same sign were considered. Although produced at a rate several times larger than that of coincidental $W$ events, less than one such event is expected to pass the analysis selction, so it is also safe to ignore this process.

\begin{figure}[htbp!]
\centering
\subfigure[]{\includegraphics[width=0.45\textwidth]{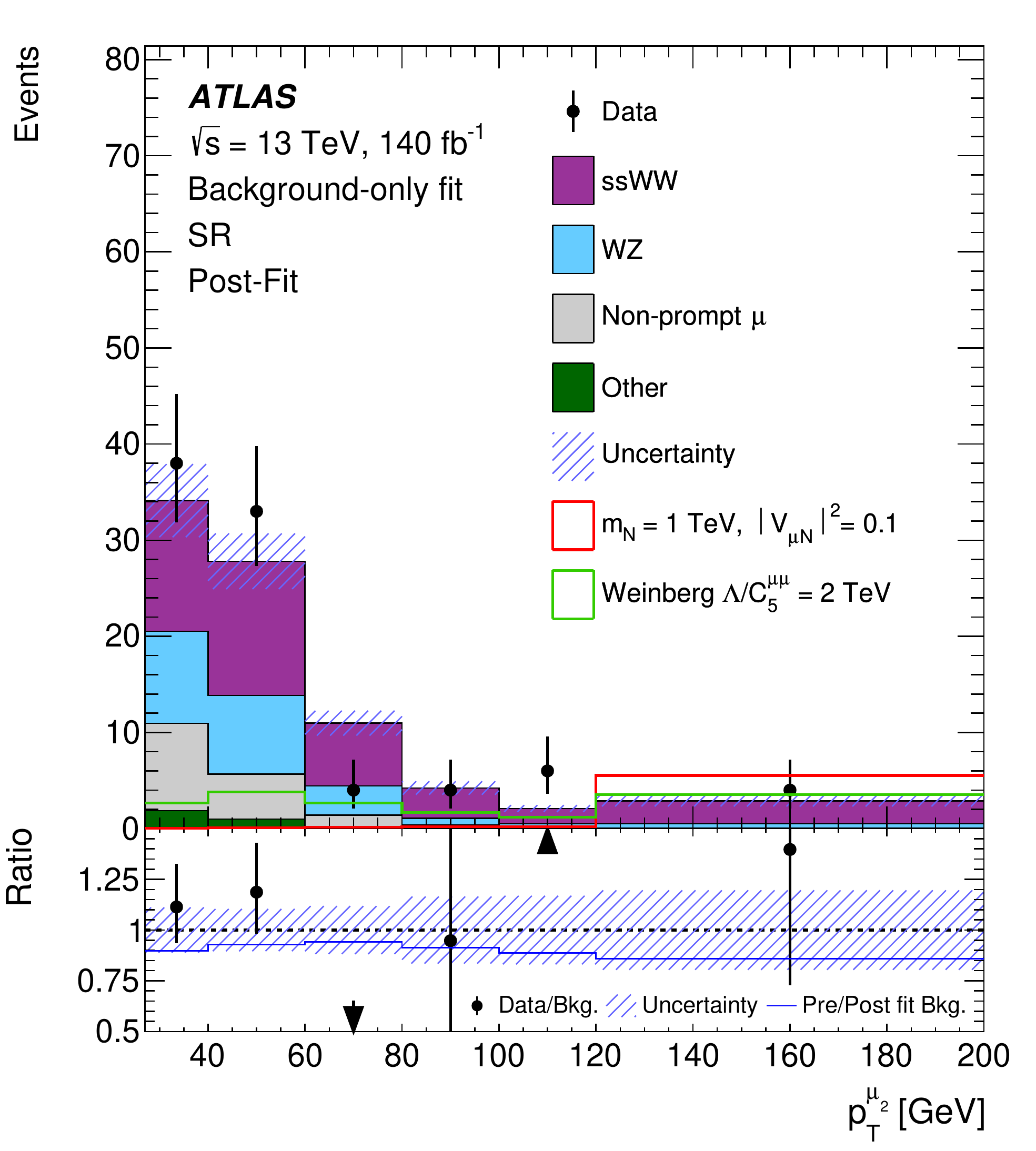}}
\\    \subfigure[]{\includegraphics[width=0.45\textwidth]{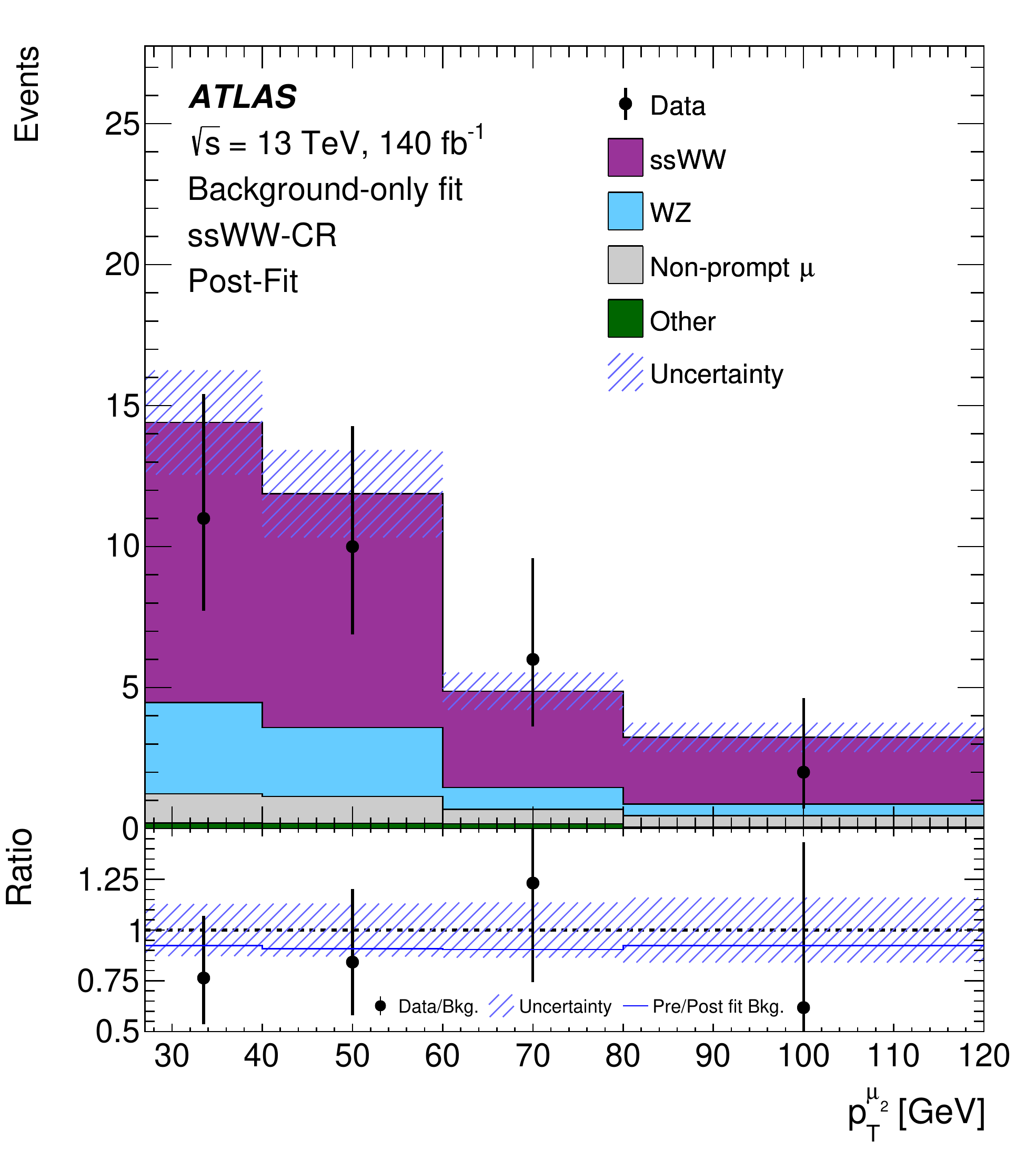}}
\subfigure[]{\includegraphics[width=0.45\textwidth]{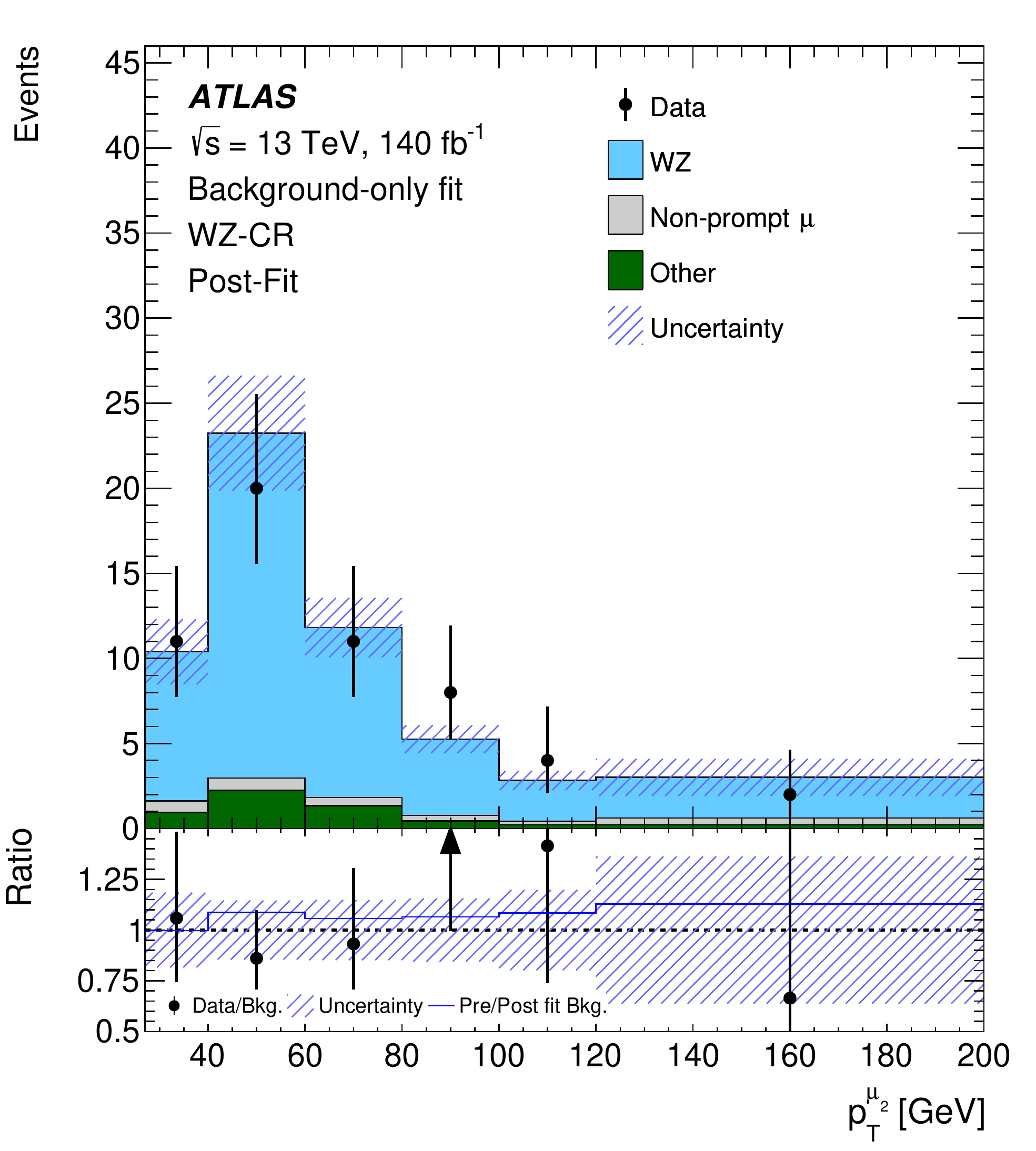}}
\caption{
Comparison of the \pTmu distributions in data and the post-fit prediction
under the background-only hypothesis in the different analysis regions:
(a) SR, (b) ssWW-CR, and (c) WZ-CR.
The shaded area surrounding the background expectation represents
the total uncertainty from the post-fit computation. The ratio of the observed yields to the prediction of the background model post-fit is shown by the points in the bottom plot and the change in the prediction relative to the pre-fit yields is shown by the blue line. The \enquote{Other} contribution consists mainly of $ZZ$ and \tZq events.
Two benchmark signal samples are overlaid on top of the total background prediction in~(a), and were obtained by using their theoretical cross sections. The last bin in (a) and (c) includes the overflow bin.
}
\label{fig:postfit_regions_Sec5}
\end{figure}


 
\section{Results}
 
This search is conducted using the binned \pTmu distributions.
A combined profile likelihood fit is used in the SR and all CRs
to obtain the final background model, as shown in Figure~\ref{fig:postfit_regions_Sec5},
and make further statistical inferences regarding the presence of a signal. The likelihood function depends on the free-floating signal normalisation,
the free-floating SM $W^{\pm}W^{\pm}jj$ and $WZ$ normalisations, and a set of nuisance parameters encoding the effect of systematic uncertainties.
The normalisation corrections obtained by the fit (defined as the ratio between post-fit and pre-fit predictions) are $1.17\pm 0.23$ for SM $W^{\pm}W^{\pm}jj$ and $0.92\pm 0.16$ for $WZ$ backgrounds. The nuisance parameters, implemented as Gaussian constraints, allow variations of the expectations for signal and background
according to the corresponding systematic uncertainties,
whilst penalising the likelihood function for any deviation from the nominal expectations according to the specified constraints.
The fit also reduces the impact of systematic uncertainties on the search sensitivity
by exploiting the background-dominated CRs. The fit and the subsequent statistical analysis are performed for each signal hypothesis.
 
Uncertainties affecting the measurement which originate from statistical sources are considered together
with systematic uncertainties related to the detector calibration and physics modelling. Uncertainty sources related to muon reconstruction and energy calibration, jet energy calibration, $b$-jet identification, rejection of \pileup jets, and missing transverse momentum reconstruction were studied. The uncertainty in the non-prompt muon background is evaluated by propagating both the systematic and statistical uncertainties in the fake factors and in the spurious-SR non-prompt muon yields. The systematic uncertainty is dominated by the modelling of the contamination from prompt muons in the region used to extract the fake factors. The statistical uncertainty, which dominates the uncertainty in the non-prompt background, is driven by the limited size of the spurious-SR data set. Systematic uncertainties in the physics modelling of SM same-sign $WW$ and $WZ$ events are estimated by varying the factorisation (\muF) and renormalisation (\muR) scales, as well as the strong coupling constant \alphas and the PDF choice. Factorisation and renormalisation scale variations are used to account for the uncertainty due to truncated higher-order corrections. The effects of a seven-point pairwise variation of \muF and \muR, where each scale is either halved or doubled (the off-diagonal variations by $[0.5, 2]$ and $[2, 0.5]$ are not included), are computed and then combined into a single uncertainty by forming an envelope of the effects of these seven variations. The total PDF uncertainty is obtained from the standard deviation of \NNPDF[3.0nlo] variations
and the envelope of alternative PDFs (\CT[14] and \MMHT), while the effect of the \alphas uncertainty is obtained by varying the nominal $\alphas=0.118$ within its uncertainty of $\pm 0.001$. Alternative SM same-sign $WW$ ($WZ$) samples generated with \SHERPA[2.2.12] and \MGNLO interfaced to \HERWIG[7.2]~\cite{Bellm:2015jjp} (\MGNLO[2.2.2] interfaced to \PYTHIA[8.186]) are used to evaluate the residual generator dependencies. A similar scheme is applied to derive uncertainties from the $\muF / \muR$ scale variations and from the PDF/\alphas choice for the \textsc{HeavyN} and \textsc{SMWeinberg} signal samples, while other smaller backgrounds are assigned a conservative $\pm 50\%$ uncertainty correlated across analysis bins. 
The largest systematic uncertainty in the total background estimate is the uncertainty in the non-prompt-muon background estimation (8\%).
None of the considered systematic uncertainties were pulled appreciably by the fit or found to have an impact on the analysis sensitivity, which is driven by the statistical uncertainty of the data in the tail of the \pTmu distribution. The expected limits change at the per-mille level if systematic uncertainties are not considered.
 
In the absence of any significant excess above the background expectation,
upper limits on \VmuN and $|C^{\mu\mu}_5| / \Lambda$ are derived with the $\text{CL}_{\textrm{s}}$ method~\cite{Junk_1999,Read:2002hq}.
The asymptotic approximation~\cite{Cowan:2010js} is used to derive the upper limits, which agree very well with those estimated via background-only pseudo-experiments.
The observed and expected 95\% confidence level (CL) upper limits on the heavy Majorana neutrino mixing element \VmuN  are shown in Figure~\ref{fig:statistics-limitplot} as a function of $m_N$. The form of the exclusion curve follows what is expected from the bare cross section $\sigma_0$ as defined in Eq.~(\ref{eq:bareXSecT}), which rises moderately with increasing $m_N$, reaching a maximum around 700~\GeV\ for 13~\TeV\ $pp$ collisions before decreasing again due to a kinematic suppression in the $W^{\pm}W^{\pm}$ matrix element~\cite{Fuks:2020att}. The strongest exclusion is achieved at $m_N=500$~\GeV, with an observed (expected) upper limit of 0.10~(0.08) on \VmuN. For a heavy neutrino mass of 20~\TeV\ the exclusion curve reaches the mixing element's upper bound of unity. Overall, the expected limits constitute a sensitivity improvement of 15\%--20\% compared to the results in Ref.~\cite{CMS-EXO-21-003}. Over the entire mass range the observed limit is around one standard deviation weaker than the expected limit, which is due to the slight excess of data in the last two bins of the \pTmu distribution. The observed (expected) 95\% CL lower limit on the $d=5$ Weinberg operator $\Lambda/ |C_5^{\mu\mu}|$ is 3.635~(4.608)~\TeV, with a probability of $p_0= 9.3\%$ that the observed data are compatible with the background-only hypothesis. Using Eq.~(\ref{eq:dim5Mass}), this translates into an upper limit of 16.7~(13.1)~\GeV\ on the effective $\mu\mu$ Majorana neutrino mass.
 
\begin{figure}[!htbp]
\centering
\includegraphics[width=.7\textwidth]{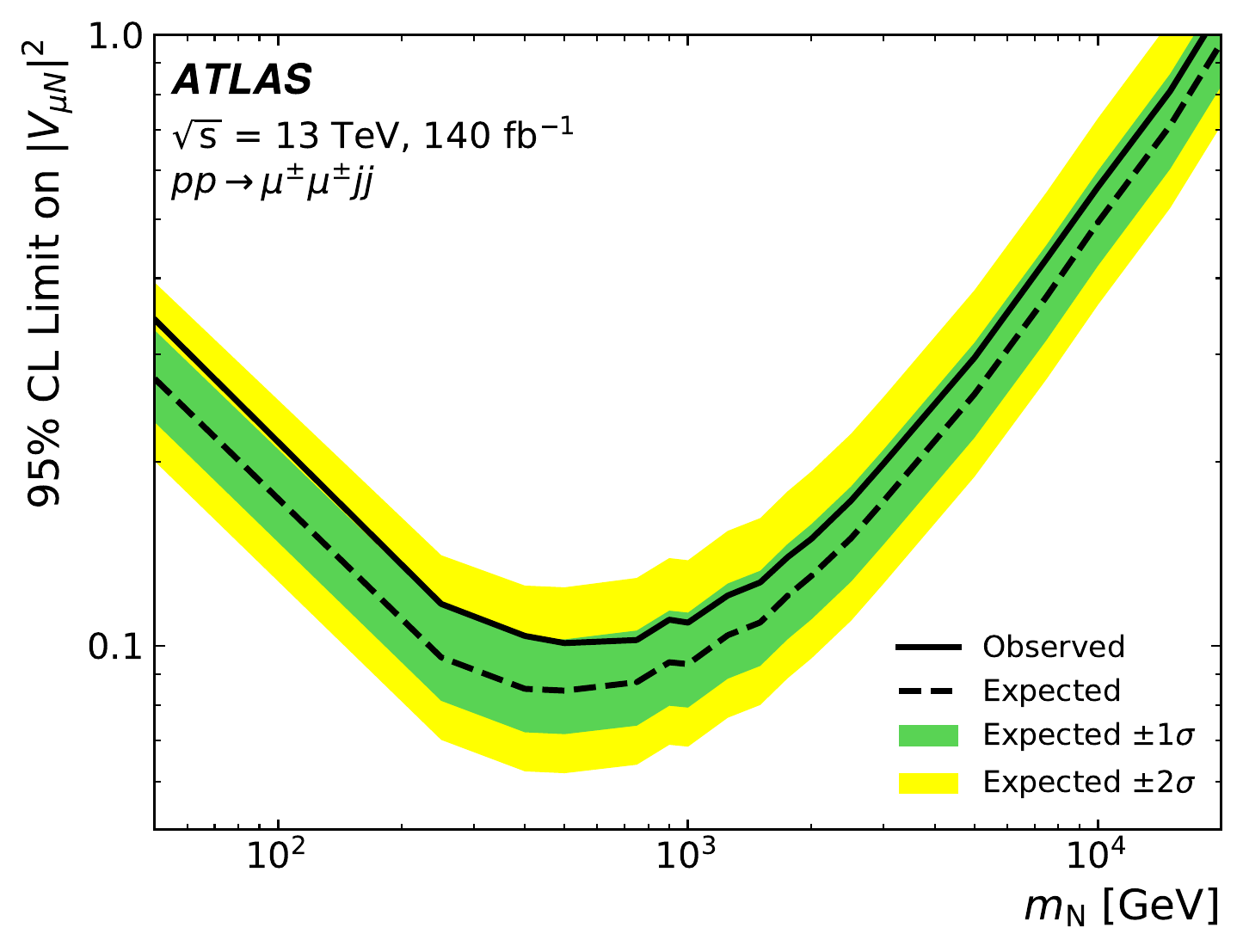}
\caption{Observed and expected 95\% CL upper limits on the heavy Majorana neutrino mixing element \VmuN  as a function of $m_N$ in the Phenomenological Type-I Seesaw model.
The one and two standard deviation bands of the expected limit are indicated in green and yellow, respectively. }
\label{fig:statistics-limitplot}
\end{figure}


 
\section{Conclusion}
 
A new search for heavy Majorana neutrinos is presented, targeting the same-sign $WW$ production mode. Since in this process the heavy neutrinos are produced in the $t$-channel, this analysis extends the reach to higher neutrino masses and complements previous searches in resonant production channels by adding valuable sensitivity for neutrino masses of a few hundred \GeV. The search is performed using 140~\ifb\ of $pp$ collision data at $\sqrt{s} = 13~\TeV$ collected by the ATLAS detector at the LHC. To preferentially capture the signature of same-sign $WW$ scattering in $pp$ collisions, events are required to have two jets with a large rapidity separation and two muons with the same electric charge. Since signal events typically have a significantly higher muon transverse momentum than those arising from SM backgrounds, the shape of the transverse momentum distribution of the subleading muon is used to discriminate between the signal and background. No significant excess was observed above the expected backgrounds, which were modelled in dedicated control regions. Upper limits on the muon-neutrino--heavy-neutrino mass-mixing matrix element \VmuN at 95\% confidence level are computed
for signals with Majorana neutrino masses $m_N$ between 50~\GeV\ and 20~\TeV\ based on a Phenomenological Type-I Seesaw model.
The strongest exclusion was achieved at $m_N=500$~\GeV, with an observed (expected) limit of 0.10~(0.08) on \VmuN.
Constraints on \VmuN are extended from previous results at $m_N$ values of the order of 1~\TeV\ in the resonant production channels to $m_N$ values of up to 20~\TeV\ in this analysis.
In addition, limits are set on the $d=5$ Weinberg operator, which translate to an observed (expected) upper limit of 16.7~(13.1)~\GeV\ on the effective $\mu\mu$ Majorana neutrino mass, which cannot be probed in nuclear decays.


\section*{Acknowledgements}


We thank CERN for the very successful operation of the LHC, as well as the
support staff from our institutions without whom ATLAS could not be
operated efficiently.
 
We acknowledge the support of
ANPCyT, Argentina;
YerPhI, Armenia;
ARC, Australia;
BMWFW and FWF, Austria;
ANAS, Azerbaijan;
CNPq and FAPESP, Brazil;
NSERC, NRC and CFI, Canada;
CERN;
ANID, Chile;
CAS, MOST and NSFC, China;
Minciencias, Colombia;
MEYS CR, Czech Republic;
DNRF and DNSRC, Denmark;
IN2P3-CNRS and CEA-DRF/IRFU, France;
SRNSFG, Georgia;
BMBF, HGF and MPG, Germany;
GSRI, Greece;
RGC and Hong Kong SAR, China;
ISF and Benoziyo Center, Israel;
INFN, Italy;
MEXT and JSPS, Japan;
CNRST, Morocco;
NWO, Netherlands;
RCN, Norway;
MEiN, Poland;
FCT, Portugal;
MNE/IFA, Romania;
MESTD, Serbia;
MSSR, Slovakia;
ARRS and MIZ\v{S}, Slovenia;
DSI/NRF, South Africa;
MICINN, Spain;
SRC and Wallenberg Foundation, Sweden;
SERI, SNSF and Cantons of Bern and Geneva, Switzerland;
MOST, Taiwan;
TENMAK, T\"urkiye;
STFC, United Kingdom;
DOE and NSF, United States of America.
In addition, individual groups and members have received support from
BCKDF, CANARIE, Compute Canada and CRC, Canada;
PRIMUS 21/SCI/017 and UNCE SCI/013, Czech Republic;
COST, ERC, ERDF, Horizon 2020 and Marie Sk{\l}odowska-Curie Actions, European Union;
Investissements d'Avenir Labex, Investissements d'Avenir Idex and ANR, France;
DFG and AvH Foundation, Germany;
Herakleitos, Thales and Aristeia programmes co-financed by EU-ESF and the Greek NSRF, Greece;
BSF-NSF and MINERVA, Israel;
Norwegian Financial Mechanism 2014-2021, Norway;
NCN and NAWA, Poland;
La Caixa Banking Foundation, CERCA Programme Generalitat de Catalunya and PROMETEO and GenT Programmes Generalitat Valenciana, Spain;
G\"{o}ran Gustafssons Stiftelse, Sweden;
The Royal Society and Leverhulme Trust, United Kingdom.
 
The crucial computing support from all WLCG partners is acknowledged gratefully, in particular from CERN, the ATLAS Tier-1 facilities at TRIUMF (Canada), NDGF (Denmark, Norway, Sweden), CC-IN2P3 (France), KIT/GridKA (Germany), INFN-CNAF (Italy), NL-T1 (Netherlands), PIC (Spain), ASGC (Taiwan), RAL (UK) and BNL (USA), the Tier-2 facilities worldwide and large non-WLCG resource providers. Major contributors of computing resources are listed in Ref.~\cite{ATL-SOFT-PUB-2021-003}.


\printbibliography
\clearpage
 
\begin{flushleft}
\hypersetup{urlcolor=black}
{\Large The ATLAS Collaboration}

\bigskip

\AtlasOrcid[0000-0002-6665-4934]{G.~Aad}$^\textrm{\scriptsize 102}$,
\AtlasOrcid[0000-0002-5888-2734]{B.~Abbott}$^\textrm{\scriptsize 120}$,
\AtlasOrcid[0000-0002-1002-1652]{K.~Abeling}$^\textrm{\scriptsize 55}$,
\AtlasOrcid[0000-0001-5763-2760]{N.J.~Abicht}$^\textrm{\scriptsize 49}$,
\AtlasOrcid[0000-0002-8496-9294]{S.H.~Abidi}$^\textrm{\scriptsize 29}$,
\AtlasOrcid[0000-0002-9987-2292]{A.~Aboulhorma}$^\textrm{\scriptsize 35e}$,
\AtlasOrcid[0000-0001-5329-6640]{H.~Abramowicz}$^\textrm{\scriptsize 151}$,
\AtlasOrcid[0000-0002-1599-2896]{H.~Abreu}$^\textrm{\scriptsize 150}$,
\AtlasOrcid[0000-0003-0403-3697]{Y.~Abulaiti}$^\textrm{\scriptsize 117}$,
\AtlasOrcid[0000-0003-0762-7204]{A.C.~Abusleme~Hoffman}$^\textrm{\scriptsize 137a}$
% ...
\bigskip
\\

% ...
$^{29}$Physics Department, Brookhaven National Laboratory, Upton NY; United States of America.\\
% ...
$^{35}$$^{(a)}$Facult\'e des Sciences Ain Chock, R\'eseau Universitaire de Physique des Hautes Energies - Universit\'e Hassan II, Casablanca;$^{(b)}$Facult\'{e} des Sciences, Universit\'{e} Ibn-Tofail, K\'{e}nitra;$^{(c)}$Facult\'e des Sciences Semlalia, Universit\'e Cadi Ayyad, LPHEA-Marrakech;$^{(d)}$LPMR, Facult\'e des Sciences, Universit\'e Mohamed Premier, Oujda;$^{(e)}$Facult\'e des sciences, Universit\'e Mohammed V, Rabat;$^{(f)}$Institute of Applied Physics, Mohammed VI Polytechnic University, Ben Guerir; Morocco.\\
% ...
$^{49}$Fakult\"{a}t Physik , Technische Universit{\"a}t Dortmund, Dortmund; Germany.\\
% ...
$^{55}$II. Physikalisches Institut, Georg-August-Universit\"{a}t G\"ottingen, G\"ottingen; Germany.\\
% ...
$^{102}$CPPM, Aix-Marseille Universit\'e, CNRS/IN2P3, Marseille; France.\\
% ...
$^{117}$Department of Physics, New York University, New York NY; United States of America.\\
% ...
$^{120}$Homer L. Dodge Department of Physics and Astronomy, University of Oklahoma, Norman OK; United States of America.\\
% ...
$^{137}$$^{(a)}$Departamento de F\'isica, Pontificia Universidad Cat\'olica de Chile, Santiago;$^{(b)}$Millennium Institute for Subatomic physics at high energy frontier (SAPHIR), Santiago;$^{(c)}$Instituto de Investigaci\'on Multidisciplinario en Ciencia y Tecnolog\'ia, y Departamento de F\'isica, Universidad de La Serena;$^{(d)}$Universidad Andres Bello, Department of Physics, Santiago;$^{(e)}$Instituto de Alta Investigaci\'on, Universidad de Tarapac\'a, Arica;$^{(f)}$Departamento de F\'isica, Universidad T\'ecnica Federico Santa Mar\'ia, Valpara\'iso; Chile.\\
% ...
$^{150}$Department of Physics, Technion, Israel Institute of Technology, Haifa; Israel.\\
$^{151}$Raymond and Beverly Sackler School of Physics and Astronomy, Tel Aviv University, Tel Aviv; Israel.\\
% ...
$^{a}$ Associated at  ¿Université catholique de Louvain¿; Belgium.\\
$^{b}$ Also Affiliated with an institute covered by a cooperation agreement with CERN.\\
$^{c}$ Also at An-Najah National University, Nablus; Palestine.\\
$^{d}$ Also at APC, Universit\'e Paris Cit\'e, CNRS/IN2P3, Paris; France.\\
$^{e}$ Also at Borough of Manhattan Community College, City University of New York, New York NY; United States of America.\\
$^{f}$ Also at Center for High Energy Physics, Peking University; China.\\
% ...

\end{flushleft}


\end{document}